\documentclass[preprint,showpacs,preprintnumbers,amsmath,amssymb,
superscriptaddress,floatfix]{revtex4}
\usepackage{graphicx}
\usepackage{amsmath}
\usepackage{bm}
\usepackage{mathrsfs}
\usepackage{color}
\usepackage{slashed}
\usepackage{dcolumn}
\usepackage{CJK}

\newcommand{\be}{\begin{equation}}
\newcommand{\ee}{\end{equation}}
\newcommand{\ba}{\begin{eqnarray}}
\newcommand{\ea}{\end{eqnarray}}
\newcommand{\nn}{\nonumber}

\newcommand{\la}{\langle}
\newcommand{\ra}{\rangle}

\begin{document}
\begin{CJK*}{GB}{gbsn}

\title{The $D\bar{D}^*$ interaction with isospin zero in an extended hidden gauge symmetry approach}

\author{Bao-Xi Sun (孙宝玺)}
\email{sunbx@bjut.edu.cn}
\affiliation{College of Applied Sciences, Beijing University
of Technology, Beijing 100124, China}
\affiliation{Department of Physics, Peking University,
Beijing 100871, China}

\author{Da-Ming Wan (万达明)}
\affiliation{College of Applied Sciences, Beijing University
of Technology, Beijing 100124, China}

\author{Si-Yu Zhao (赵思宇)}
\affiliation{College of Applied Sciences, Beijing University
of Technology, Beijing 100124, China}

\date{\today}

\begin{abstract}
The $D \bar{D}^*$ interaction via a $\rho$ or $\omega$ exchange is
constructed within an extended hidden gauge symmetry approach, where
the strange quark is replaced by the charm quark in the $SU(3)$ flavor
space.
With this $D \bar{D}^*$ interaction, a bound state slightly lower
than the $D \bar{D}^*$ threshold is generated dynamically in the
isospin zero sector by solving the Bethe-Salpeter equation in the
coupled-channel approximation, which might correspond to the
$X(3872)$ particle announced by many collaborations.
This formulism is also used to study the $B \bar{B}^*$ interaction,
and a $B \bar{B}^*$ bound state with isospin zero is generated
dynamically, which has no counterpart listed in the review of the
Particle Data Group.
Furthermore, the one-pion exchange between the $D$ meson and the
$\bar{D}^*$ is analyzed precisely, and we do not think the one-pion
exchange potential need be considered when the Bethe-Salpeter
equation is solved.
\end{abstract}


\maketitle

\section{Introduction}
\label{sect:Intro3872}

The hidden gauge symmetry approach has been shown to be a successful method to include the vector meson in the Lagrangian~\cite{Bando:1984ej,Bando:1987br,Meissner:1987ge,Harada:2003jx}. Along these lines, the pseudoscalar meson and vector meson interaction~\cite{Nagahiro:2008cv}, the vector meson and vector meson interaction~\cite{Molina:2008jw,Geng:2008gx}, the vector meson and baryon octet interaction~\cite{Oset:2009vf,Khemchandani:2011et}, and the vector meson and baryon decuplet interaction~\cite{Gonzalez:2008pv,Sarkar09} in $SU(3)$ flavor space have been studied in the coupled-channel unitary approximation. This method is extended to $SU(4)$ space when the components related to $c$ and $\bar{c}$ quarks are taken into account~\cite{Wu:2010jy,Wu:2010vk,Xiao:2013yca}.
In the past few years, more and more XYZ states have been discovered, and it has become necessary to include the $c$ and $b$ quark components in the effective Lagrangian when the interaction of hadrons is investigated.
However, since
mesons composed of $c$ and $b$ quarks are much heavier than mesons composed of light quarks, the exchange of heavier mesons is extremely suppressed, and the mesons which consist completely of light quarks, such as pions, and $\rho$ and $\omega$ mesons, play a dominant role in the interactions of hadrons.

The $c$ and $b$ quarks usually act as spectators in the interactions of hadrons. Thus, strange quarks can be replaced by $c$ or $b$ quarks in the process of strangeness zero, and then the interactions of hadrons composed of heavier flavor quarks can be discussed in the $SU(3)$ subspace of $u$, $d$ and $c(b)$ quark components. Many studies have been done on this topic~\cite{Liang:2014kra,Uchino:2015uha,Liang:2014eba}, and it should especially be stressed that this replacement is used in the study of the generation of charm-beauty bound states of $B(B^*)D(D^*)$ and $B(B^*)\bar{D}(\bar{D}^*)$ interactions~\cite{Sakai:2017avl}. It is clear that the model becomes much simpler than those used in Refs.~\cite{Wu:2010jy,Wu:2010vk,Xiao:2013yca}, where the $SU(4)$ hidden gauge symmetry approach is discussed in detail.

The $X(3872)$ state was first observed by the Belle Collaboration in 2003~\cite{Choi:2003ue}, and then confirmed by many experimental collaborations. Finally, a mass of $3871.69\pm0.17$~MeV~\cite{PDG2016} and a decay width $<1.2$~MeV~\cite{Choi:2011fc} are given by fitting the experimental data, which is extremely close to the $D\bar{D}^*$ threshold.
A lot of theoretical research work has been done on the properties of $X(3872)$. Some people suppose $X(3872)$ to be a $D \bar{D}^{*}/\bar{D} D^{*}$ bound state since its mass is very close to the $D \bar{D}^{*}$ threshold~\cite{Zhq20,Zhq21,Zhq22,Zhq23}. $X(3872)$ is also described as a virtual state of $D \bar{D}^{*}/\bar{D} D^{*}$~\cite{Zhq24,Kang:2016jxw}, a tetraquark~\cite{Zhq25,FernandezCarames:2009zz,Carames:2012th}, a hybrid state~\cite{Zhq26} or a mixture of a charmonium $\chi_{c1}(2P)$ with a $D \bar{D}^{*}/\bar{D} D^{*}$ component~\cite{Zhq27,Zhq28}. Moreover, the $X(3872)$ state is studied by using the pole counting rule method~\cite{Zhq341,Zhq342}, and it is found that two nearby poles are necessary to describe the experimental data~\cite{Zhq33,Meng:2014ota}.

In the present work, we will replace the strange quark by the charm quark in the $SU(3)$ hidden gauge symmetry approach, and then study the $D\bar{D}^*$ interaction in the coupled-channel unitary approximation by solving the Bethe-Salpeter equation. Consequently, the $X(3872)$ state is generated dynamically when the $\rho$ and $\omega$ exchanges between $D$ and $\bar{D}^*$ mesons are taken into account.

One-pion exchange between $D$ and $\bar{D}^*$ mesons at the
$D\bar{D}^*$ threshold is addressed specially. Since the mass of the
$\bar{D}^*$ meson is about one pion mass larger than the mass of the
$D$ meson, the intermediate pion might be regarded as a $real$
particle at the $D\bar{D}^*$ threshold, therefore the behavior of
the $DD^*$ interaction through one-pion exchange is interesting.
However, although it is divergent at the $D\bar{D}^*$ threshold, the
one-pion exchange potential of $D\bar{D}^*$ becomes weaker when the
total energy of the system departs from the $D\bar{D}^*$ threshold.
When the hidden gauge symmetry approach is considered, the $\rho$
and $\omega$ meson exchange between $D$ and $\bar{D}^*$ mesons is
dominant, and thus the one-pion exchange potential is neglected in
the present work.

In addition, this model is extended to study the $B\bar{B}^*$
interaction in the isospin zero sector by replacing the $c$ quark
with a $b$ quark, and a new bound state is predicted, which is not
listed in the review of the Particle Data Group~(PDG)~\cite{PDG2016}.

This article is organized as follows. The formulism is described in
Section~\ref{sect:formalism}, and then the implementation of
unitarity is discussed in Section~\ref{sect:unitarity}, where the
contribution from the longitudinal part of the vector meson
propagator in the loop function of the Bethe-Salpeter equation is
taken into account. The one-pion exchange potential of $D\bar{D}^*$
is analyzed in Section~\ref{sect:uchannel}. The calculation results
on the $D\bar{D}^*$ and $B\bar{B}^*$ interactions are presented in
Section~\ref{sect:results3872}. Finally, a summary is given in
Section~\ref{sect:summary3872}.

\section{Formalism}
\label{sect:formalism}

The hidden gauge symmetry approach is successful when vector mesons are involved in the Lagrangian, where vector mesons are treated as gauge bosons of the $SU(3)$ local gauge symmetry breaking spontaneously~\cite{Bando:1984ej, Bando:1987br,Meissner:1987ge,Harada:2003jx,Nagahiro:2008cv}. This formalism can be extended to study the interaction of the $D$ meson and the $\bar{D}^*$ meson by replacing the $s$ and $\bar{s}$ quarks with $c$ and $\bar{c}$ quarks, respectively.

\begin{figure}[!htb]
\centerline{
\includegraphics[width = 0.65\linewidth]{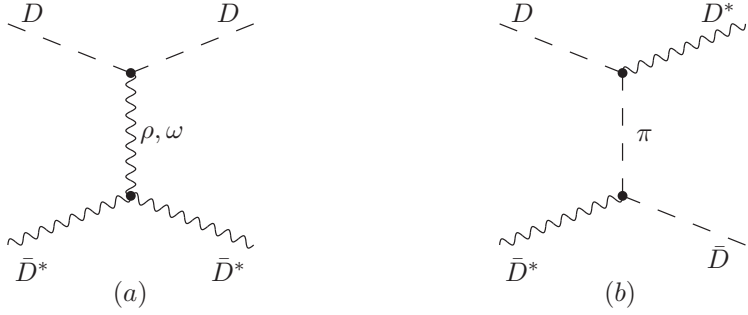}
}
\caption{The interactions of the $D$ and $\bar{D}^*$ mesons. (a)~Vector meson exchange, (b)~One-pion exchange. }
\label{fig:exchange}
\end{figure}

In the hidden gauge symmetry approach, the $D \bar{D}^*$ interaction would proceed through the exchange of a vector meson, as depicted in Fig.~\ref{fig:exchange}(a). Since the vector propagator contributes a factor of $1/M_V^2$ if the momentum transfer between the $D$ meson and the $\bar{D}^*$ meson can be neglected, the exchange of $\rho$ and $\omega$ mesons is dominant, while the possible exchange of heavier vector mesons is suppressed.

The $DD\rho$ and $DD\omega$ couplings can be obtained with the Lagrangian
\be
\label{eq:Lvpp}
{\cal L}=-i g \la V_\mu [P, \partial^\mu P ] \ra,
\ee
where
\be
g=\frac{M_V}{2f_\pi},
\ee
with $f_\pi=93$MeV the pion decay constant and $M_V$ the mass of the $\rho$ meson.

The matrices of vector mesons and pseudoscalar mesons take the form of
\be
V_\mu=\left(\begin{array}{ccc}
 \frac{\omega}{\sqrt{2}}+\frac{\rho^0}{\sqrt{2}} & \rho^+ & \bar{D}^{*0}  \\
 \rho^-                                          & \frac{\omega}{\sqrt{2}}-\frac{\rho^0}{\sqrt{2}} & D^{*-} \\
  D^{*0}                                         & D^{*+}                                          &  0
                                               \end{array}
                                          \right),
\ee
and
\be
P=\left(\begin{array}{ccc}
 \frac{\pi^0}{\sqrt{2}} & \pi^+ & \bar{D}^{0}  \\
 \pi^- & -\frac{\pi^0}{\sqrt{2}} &      D^{-} \\
D^{0} & D^+ &  0
                                               \end{array}
                                          \right),
\ee
respectively, where only the relevant mesons are enumerated.

The Lagrangian density of vector mesons can be written as
\be
\label{eq:Lagrangianvectormeson}
{\cal L}_{V} =-\frac{1}{4} \la V_{\mu \nu} V^{\mu \nu} \ra,
\ee
with
\be
V_{\mu \nu}= \partial_\mu V_\nu - \partial_\nu V_\mu -ig [V_\mu, V_\nu].
\ee
According to Eq.~(\ref{eq:Lagrangianvectormeson}), we can derive the $D^*D^*\rho$ and $D^* D^* \omega$ couplings from the interaction Lagrangian
\be
{\cal L}_{VVV}=ig \la (\partial_\mu V_\nu - \partial_\nu V_\mu) V^\mu V^\nu       \ra.
\ee
Since the mass of the $\omega$ meson $m_\omega=782$~MeV is similar to that of the $\rho$ meson $m_\rho=770$~MeV, we suppose $M_V\approx m_\rho \approx m_\omega$, then the potential of the $D$ meson and $\bar{D}^*$ meson is simplified as
\be
\label{eq:Vdddd3872}
V_{ij}=C_{ij}\frac{1}{f_\pi^2} [(k_1+k_2)\cdot (p_1+p_2)]\varepsilon \cdot \varepsilon^*,
\ee
with $\varepsilon$ and $\varepsilon^*$ the polarization vectors of the initial and final
vector mesons, and $k_1(p_1)$ and $k_2(p_2)$ the momenta of the initial and final $D(\bar{D}^*)$ mesons, respectively.
The coefficients $C_{ij}$ in the different channels are shown in Table~\ref{table:coef_DDstar}.

\begin{table}[htbp]
\begin{tabular}{c|cccc}
\hline\hline
 $C_{ij}$          &$D^+ D^{*-}$ & $D^0 \bar{D}^{*0}$ & $\bar{D}^{0} D^{*0}$ & $D^- D^{*+}$    \\
\hline
 $D^+ D^{*-}$      &$\frac{1}{4}$ & $\frac{1}{4}$ & $0$ & $0$    \\
 $D^0 \bar{D}^{*0}$&$\frac{1}{4}$   & $\frac{1}{4}$ & $0$ & $0$    \\

$\bar{D}^{0}D^{*0}$&$0$ & $0$   & $\frac{1}{4}$ & $\frac{1}{4}$    \\

$D^- D^{*+}$       &$0$   & $0$   & $\frac{1}{4}$   & $\frac{1}{4}$    \\
 \hline \hline
\end{tabular}
\caption{The coefficients $C_{ij}$ in the $D$ and $\bar{D}^*$ interaction, $C_{ji}=C_{ij}$.}
\label{table:coef_DDstar}
\end{table}

The $D \bar{D}^*$ pair  with isospin $I=0$ takes the form of
\be
\label{eq:isospinfunc}
| D \bar{D}^*, I=0 \ra = \frac{1}{\sqrt{4}} \left(|D^+ D^{*-} \ra + |D^0 \bar{D}^{*0}\ra
-|\bar{D}^{0} D^{*0}\ra - |D^- D^{*+}\ra
\right),
\ee
where the C-parity of the $D \bar{D}^*$ pair is assumed to be positive.

According to Eqs.~(\ref{eq:Vdddd3872}) and ~(\ref{eq:isospinfunc}), the potential of $D \bar{D}^*$ with isospin $I=0$ can be written as
\be
\label{eq:Vdddd38722}
V^{t}_{D \bar{D}^* \rightarrow D \bar{D}^*}=\frac{1}{2}\frac{1}{f_\pi^2} [(k_1+k_2)\cdot (p_1+p_2)]\varepsilon \cdot \varepsilon^*.
\ee

According to Ref.~\cite{Sun201704}, the kernel $\tilde{V}^{t}_{D \bar{D}^* \rightarrow D \bar{D}^*}$ can be obtained from the potential form in Eq.~(\ref{eq:Vdddd38722}) when the Bethe-Salpeter equation is solved, i.e.,
\be
\label{eq:tildeV1}
\tilde{V}^{t}_{D \bar{D}^* \rightarrow D \bar{D}^*}=\frac{1}{2}\frac{1}{f_\pi^2} [(k_1+k_2)\cdot (p_1+p_2)],
\ee
where the $\varepsilon \cdot \varepsilon^*$ structure has been eliminated.

Actually, the kernel in Eq.~(\ref{eq:tildeV1}) can be written as
\ba
\label{eq:tildeV1222222}
\tilde{V}^{t}_{D \bar{D}^* \rightarrow D \bar{D}^*}&=&\frac{1}{2}\frac{1}{f_\pi^2} (s-u) \nn \\
&=&\frac{1}{2}\frac{1}{f_\pi^2} \left( 2s+t-2(M_D^2+M_{D^*}^2)  \right),
\ea
where the Mandelstam variables $s=(p_1+k_1)^2$, $t=(k_2-k_1)^2$ and $u=(p_2-k_1)^2$. In the derivation of Eq.~(\ref{eq:tildeV1}), we have neglected the momentum transfer $q=k_2-k_1$ compared to the mass of the vector meson $M_V$, which would be a good approximation for the interaction relatively
close to threshold where bound states or resonances are searched for,
i.e., $t=(k_2-k_1)^2=0$ is assumed in the approximation.
Thus the kernel in Eq.~(\ref{eq:tildeV1222222}) is only a function of
the Mandelstam variables $s$, which is the square of the total energy in
the center of mass frame.
\ba
\label{eq:tildeV133333}
\tilde{V}^{t}_{D \bar{D}^* \rightarrow D \bar{D}^*}
&=&\frac{1}{f_\pi^2} \left( s-M_D^2-M_{D^*}^2  \right).
\ea

\section{Implementation of unitarity}
\label{sect:unitarity}

In the coupled-channel unitary approach, the unitarity can be implemented into the $D\bar{D}^*$ interaction by solving the Bethe-Salpeter equation:
\ba
\tilde{T}
&=&[I-\tilde{V}\tilde{G}]^{-1} \tilde{V}, \label{eq:Bethe}
\ea
where $\tilde{V}$ is the kernel of the $D\bar{D}^*$ interaction
provided by Eq.~(\ref{eq:tildeV1}), and $\tilde{G}$ is the
$D\bar{D}^*$ loop function. The loop function is logarithmically
divergent and thus is calculated with a three-momentum
cutoff~\cite{Ramos97,Oller99}, or by means of dimensional
regularization~\cite{Oller2001}. Recently, a loop function of a
pseudoscalar meson and a vector meson is derived in the dimensional
regularization scheme, where the contribution of the longitudinal
part of the vector meson propagator is taken into account in
 Ref.~\cite{Sun201704}.
In the present work, this formula of the loop function will be applied to the $D\bar{D}^*$ interaction in the hidden gauge symmetry approach.

The loop function can be written as
\be
\label{eq:20170820}
\tilde{G}(s)=-\left( G_{D^*D}(s)+\frac{1}{M_{D^*}^2}H^{00}_{D^*D}(s)+\frac{s}{4M_{D^*}^2}H^{11}_{D^*D}(s)\right),
\ee
where $G_{D^*D}(s)$ is the original form of the loop function in Ref.~\cite{Oller2001}, while
the terms related to $H^{00}_{D^*D}(s)$ and $H^{11}_{D^*D}(s)$ stem
from the longitudinal part of the vector meson propagator, and their
analytical forms can be found in the appendix of
Ref.~\cite{Sun201704}.

\section{One-pion exchange}
\label{sect:uchannel}

From the Lagrangian in Eq.~(\ref{eq:Lvpp}), we can obtain the interaction Lagrangian for the $D^*D\pi$ coupling, which can be written as
\ba
{\cal L}_{D^*D\pi}&=&-\frac{ig}{\sqrt{2}}
[ \bar{D}^{*0}_\mu(D^0 \partial^\mu \pi^0 - \partial^\mu D^0 \pi^0)
+\sqrt{2}\bar{D}^{*0}_\mu(D^+ \partial^\mu \pi^{-} - \partial^\mu D^+ \pi^{-}) \nn \\
&+&\sqrt{2}{D}^{*0}_\mu (\pi^+ \partial^\mu D^- - \partial^\mu \pi^+ D^-)
+{D}^{*0}_\mu (\pi^0 \partial^\mu \bar{D}^0 - \partial^\mu \pi^0 \bar{D}^0  )  \nn \\
&+&\sqrt{2} D^{*-}_\mu (D^0 \partial^\mu \pi^+ - \partial^\mu D^0 \pi^+ )
-D^{*-}_\mu ( D^+ \partial^\mu \pi^0 - \partial^\mu D^+ \pi^0) \nn \\
&+&\sqrt{2} D^{*+}_\mu (\pi^{-} \partial^\mu \bar{D}^0 - \partial^\mu \pi^{-} \bar{D}^0 )
- D^{*+}_\mu (\pi^0 \partial^\mu D^{-}-\partial^\mu \pi^0 D^-)].
\ea
Therefore, the one-pion exchange potential of the $D$ and $\bar{D}^*$ mesons is obtained as
\be
\label{eq:uchannel}
V_{ij}^u=D_{ij}g^2  (q - k_1 )\cdot \varepsilon^*  \frac{1}{q^2-m_\pi^2} (q-k_2) \cdot \varepsilon,
\ee
as shown in Fig.~\ref{fig:exchange}(b).
The coefficients $D_{ij}$ in Eq.~(\ref{eq:uchannel}) for different channels are listed in Table \ref{table:coef_DDstar_u_channel}. According to Eq.~(\ref{eq:isospinfunc}), the one-pion exchange potential of the $D$ and $\bar{D}^*$ mesons in the sector of isospin $I=0$ can be written as
\ba
\label{eq:uchannelisospin}
V_{D\bar{D}^* \rightarrow D\bar{D}^*}^u
&=&6 g^2  q \cdot \varepsilon^*  \frac{1}{q^2-m_\pi^2} q \cdot \varepsilon,
\ea
where $q=p_2-k_1=p_1-k_2$, $p_1 \cdot \varepsilon=0$ and $p_2 \cdot \varepsilon^*=0$ are used in the derivation.

\begin{table}[htbp]
\begin{tabular}{c|cccc}
\hline\hline
 $D_{ij}$          &$D^+ D^{*-}$ & $D^0 \bar{D}^{*0}$ & $\bar{D}^{0} D^{*0}$ & $D^- D^{*+}$    \\
\hline
 $D^+ D^{*-}$      &$0$ & $0$ & $-1$ & $-\frac{1}{2}$    \\
 $D^0 \bar{D}^{*0}$&$0$   & $0$ & $-\frac{1}{2}$ & $-1$    \\

$\bar{D}^{0}D^{*0}$&$-1$ & $-\frac{1}{2}$   & $0$ & $0$    \\

$D^- D^{*+}$       &$-\frac{1}{2}$   & $-1$   & $0$   & $0$    \\
 \hline \hline
\end{tabular}
\caption{The coefficients $D_{ij}$ in the one-pion exchange potential of the $D$ and $\bar{D}^*$ interaction, $D_{ji}=D_{ij}$.}
\label{table:coef_DDstar_u_channel}
\end{table}

Since the $\bar{D}^*$ meson mass is about one pion mass larger than that of the $D$ meson, $M_{D*}-M_D\approx m_\pi$, the intermediate pion can be regarded as a $real$ particle approximately at the threshold of $D\bar{D}^*$, i.e., $q_0^2\approx \vec{q}^2+m_\pi^2$. The denominator in the one-pion exchange potential of the $D$ and $\bar{D}^*$ mesons in Eq.~(\ref{eq:uchannelisospin}) can be written as
\ba
\label{eq:q2m2}
 &&q^2-m_\pi^2 \nn \\
&=&q_0^2-\vec{q}^2-m_\pi^2 \nn \\
&=&[\sqrt{\vec{q}^2+m_\pi^2}]^2-\vec{q}^2-m_\pi^2 \nn \\
&=&\left[m_\pi\sqrt{1+\frac{\vec{q}^2}{m_\pi^2}}\right]^2-\vec{q}^2-m_\pi^2 \nn \\
&\sim&\left[m_\pi \left(1+\frac{\vec{q}^2}{2 m_\pi^2} \right)\right]^2-\vec{q}^2-m_\pi^2 \nn \\
&\sim&\frac{|\vec{q}|^4}{(2m_\pi)^2},
\ea
approximately.
However, the zero component of the polarization vector of the
$\bar{D}^*$ meson is in inverse proportion to the $\bar{D}^*$ meson
mass, and thus can be neglected in the calculation, so we have \be
\label{eq:qdote} q \cdot \varepsilon^* \sim |\vec{q}|, \ee and \be
\label{eq:qdotepr} q \cdot \varepsilon \sim |\vec{q}|. \ee

According to Eqs.~(\ref{eq:q2m2}), ~(\ref{eq:qdote})
and~(\ref{eq:qdotepr}), although the one-pion exchange potential of
the $D$ and $\bar{D}^*$ mesons is divergent at the $D\bar{D}^*$
threshold,
\be V_{D\bar{D}^* \rightarrow D\bar{D}^*}^u\sim
\frac{1}{|\vec{q}|^2}, \ee it can be neglected when the total energy
of the system is far away from the $D\bar{D}^*$ threshold.

\begin{figure}[!htb]
\centerline{
\includegraphics[width = 0.65\linewidth]{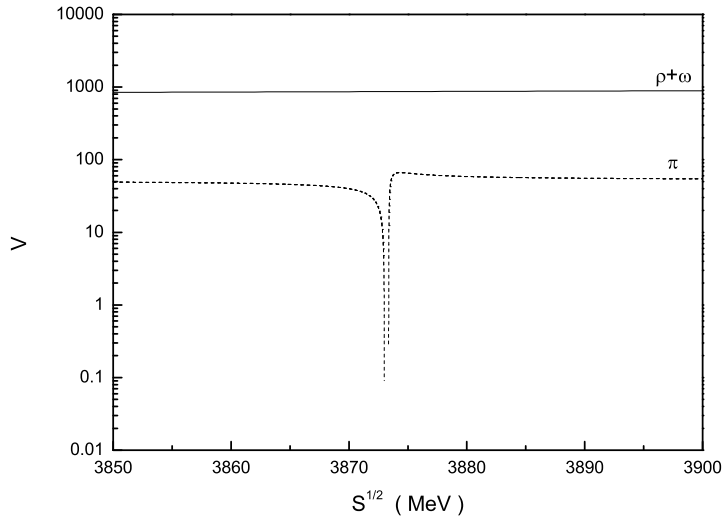}
}
\caption{The vector meson exchange potential of $D\bar{D}^*$ in
Eq.~(\ref{eq:tildeV133333}) (solid line) and the one-pion exchange
potential of $D\bar{D}^*$ in Eq.~(11) of Ref.~\cite{Zhq20} (dashed line)
as functions of the total energy of the system $\sqrt{s}$. }
\label{fig:Vrhoomepi}
\end{figure}

In Ref.~\cite{Zhq20}, the one-pion exchange potential of the $D$ and
$\bar{D}^*$ mesons is assumed to be dominant in the generation of
the $X(3872)$ state. The formula of the one-pion exchange potential is
given explicitly in the second term in Eq.~(11) of
Ref.~\cite{Zhq20}, which is relevant to the external three-momentum
in the center-of-mass frame. The potentials of the $D$ and
$\bar{D}^*$ mesons as functions of the total energy of the system
$\sqrt{s}$ are depicted in Fig.~\ref{fig:Vrhoomepi}, and it can be
found that the vector meson exchange potential is more important
than the one-pion exchange potential of the $D$ and $\bar{D}^*$
mesons if the hidden gauge symmetry is taken into account.
Therefore, the one-pion exchange potential of the $D$ and
$\bar{D}^*$ mesons is neglected in the present work.

\section{Results}
\label{sect:results3872}

In Ref.~\cite{Garzon:2013uwa}, the $D \bar{D}^*$ interaction is
studied in the $SU(4)$ flavor space, and an intermediate $J/\psi$
exchange in the kernel is taken into account besides the $\rho$ and
$\omega$ exchanges. Actually, the $J/\psi$ particle is heavier
than the $\rho$ and $\omega$ mesons, and the $D \bar{D}^*$
interaction via a $J/\psi$ exchange can be neglected in the
calculation. Moreover, we suppose that the pion decay constant
$f_\pi=93$~MeV in the $D \bar{D}^*$ potential in
Eq.~(\ref{eq:Vdddd3872}). However, the $f_\pi^2$ is replaced with
$f_i f_j$ in the potential in Eq.~(4) of Ref.~\cite{Garzon:2013uwa},
related to the initial and final particles, respectively.
In the $D \bar{D}^* \rightarrow D \bar{D}^*$ process, both $f_i$ and
$f_j$ take the value of the decay constant of the $D$ meson, i.e.,
$f_i=f_j=f_D=165$~MeV.

Five channels of $\frac{1}{\sqrt{2}}(\bar{K}^{*-}K^+ -c.c.)$,
$\frac{1}{\sqrt{2}}(\bar{K}^{*0}K^0 -c.c.)$,
$\frac{1}{\sqrt{2}}({D}^{*+}D^- -c.c.)$,
$\frac{1}{\sqrt{2}}({D}^{*0}\bar{D}^0 -c.c.)$ and
$\frac{1}{\sqrt{2}}({D}_s^{*+}D_s^- -c.c.)$ are discussed in
Ref.~\cite{Garzon:2013uwa}.
A potential is given by \be V_{ij}(s,t,u)=\frac{\xi_{ij}}{4 f_i f_j}
(s-u) \vec{\epsilon} \cdot \vec{\epsilon}^*, \ee where $\xi_{ij}$
denotes the coefficient between these channels, and $\vec{\epsilon}$
and $\vec{\epsilon}^*$ are 3-dimensional polarization vectors of the
initial and final vector mesons, respectively.
When the $J/\psi$ exchange is neglected, the coefficients $\xi_{ij}$
for the $\frac{1}{\sqrt{2}}({D}^{*+}D^- -c.c.)$ and
$\frac{1}{\sqrt{2}}({D}^{*0}\bar{D}^0 -c.c.)$ channels can be
obtained from the values listed in Table~\ref{table:coef_DDstar}. It
is apparent that $\vec{\epsilon} \cdot \vec{\epsilon}^*$ is supposed
to be $-1$ in Ref.~\cite{Garzon:2013uwa}, and thus the coefficients
in these two channels take negative values in Eq.~(5) of
Ref.~\cite{Garzon:2013uwa}.

The $\bar{K}^{*}K$ threshold is far lower than the energy region
where the $X(3872)$ is detected. Thus the
$\frac{1}{\sqrt{2}}(\bar{K}^{*-}K^+ -c.c.)$ and
$\frac{1}{\sqrt{2}}(\bar{K}^{*0}K^0 -c.c.)$ channels can be excluded
when the generation of the $X(3872)$ particle is discussed.
Moreover, it should be emphasized that the
$\frac{1}{\sqrt{2}}({D}_s^{*+}D_s^- -c.c.)$ channel only contributes
about $0.016$ of the probability in the wave function of the
$X(3872)$ particle, as discussed in Ref.~\cite{Garzon:2013uwa},
so the $\frac{1}{\sqrt{2}}({D}^{*+}D^- -c.c.)$ and
$\frac{1}{\sqrt{2}}({D}^{*0}\bar{D}^0 -c.c.)$ channels play an
important role in the generation of the $X(3872)$ particle.
Therefore, it is reasonable that only the $D \bar{D}^*$ interaction
is taken into account in the present work.

\begin{figure}[!htb]
\centerline{
\includegraphics[width = 0.65\linewidth]{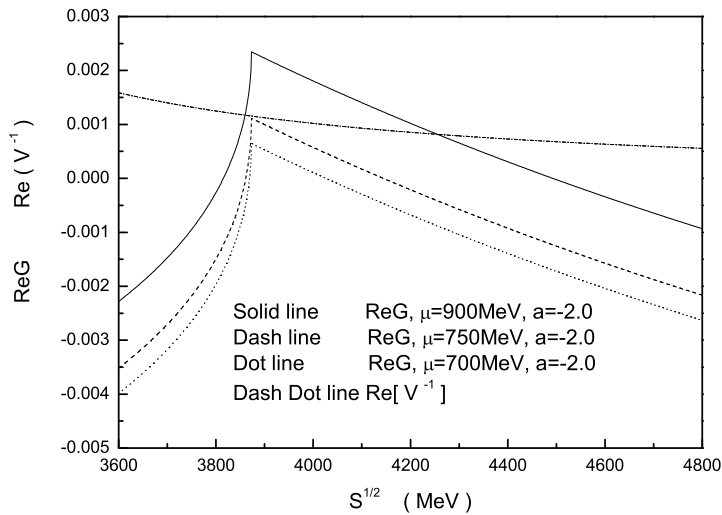}
}
\caption{The inverse of the $D\bar{D}^*$ potential in
Eq.~(\ref{eq:tildeV133333}) and the real part of the loop function
in Eq.~(\ref{eq:20170820}) with different values of the
regularization scale $\mu$ as functions of the total energy of the
system $\sqrt{s}$, where the subtraction constant $a=-2$ is fixed.
The solid, dashed and dotted lines stand for the real
part of the loop function with $\mu=900$~MeV, $\mu=750$~MeV and
$\mu=700$~MeV, respectively, while the dash-dotted line denotes the
inverse of the $D\bar{D}^*$ potential. }
\label{fig:tl19900780700}
\end{figure}

The resonance state of $D\bar{D}^*$ corresponds to the condition
\be
\label{eq:polecondition}
det(I-\tilde{V}\tilde{G})=0.
\ee
In a single channel, Eq.~(\ref{eq:polecondition}) leads to poles in
the $\tilde{T}$ amplitude when $\tilde{V}^{-1}=\tilde{G}$.
Figure~\ref{fig:tl19900780700} shows the real parts of the loop function
$\tilde{G}$ of $D\bar{D}^*$ with different values of the
regularization scale $\mu$ in Eq.~(\ref{eq:20170820}) as functions
of the total energy of the system $\sqrt{s}$ in the center-of-mass
frame. The inverse of the kernel $\tilde{V}$
in Eq.~(\ref{eq:tildeV133333}) is also shown.
The real part of the loop function $\tilde{G}$ is less
than the value of $\tilde{V}^{-1}$ when the regularization scale
$\nu<750$~MeV with $a=-2$ fixed. Therefore, no resonance state is
generated dynamically in the $D\bar{D}^*$ channel with isospin zero
even if a peak of the $\tilde{T}$ amplitude is detected on the
complex energy plane of $\sqrt{s}$.
A pole of the $\tilde{T}$ amplitude appears at
$3872.62+i0.00$~MeV in the complex energy plane of $\sqrt{s}$ if the
value of the regularization scale is set to be $\mu=800$~MeV with the
subtraction constant $a=-2$ fixed, which is consistent with the
experimental data for the $X(3872)$ particle. The real part of the
pole position is about $1\sim2$~MeV lower than the $D\bar{D}^*$
threshold, and thus the $X(3872)$ particle can be regarded as a
$D\bar{D}^*$ bound state.

If the longitudinal part of the vector propagator in the loop
function $\tilde{G}$ is excluded, a bound state can also be
generated in the corresponding energy region by adjusting values of
the regularization scale $\mu$ and the subtraction constant $a$. In
this case, a pole of the $\tilde{T}$ amplitude is detected at
$3871.69+i0.00$~MeV on the complex energy plane of $\sqrt{s}$ with
$\mu=813$~MeV and $a=-2$.

\begin{figure}[!htb]
\centerline{
\includegraphics[width = 0.65\linewidth]{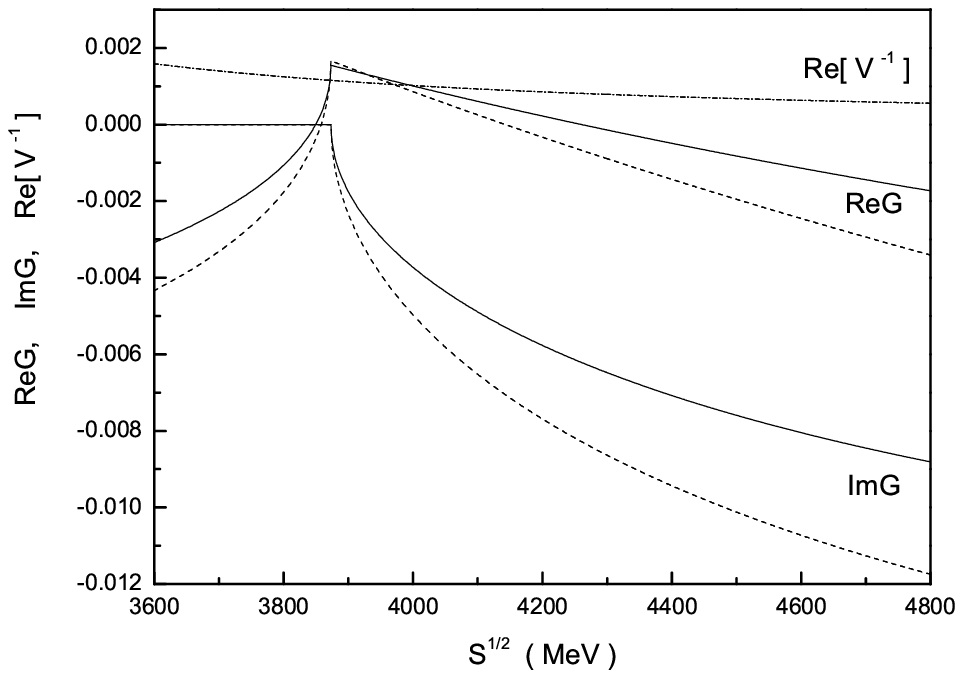}
} \caption{The inverse of the $D\bar{D}^*$ potential in
Eq.~(\ref{eq:tildeV133333}) and the real and imaginary parts of the
loop function as functions of the total energy of the system
$\sqrt{s}$. The solid lines show the case where both the
transverse and longitudinal parts of the vector meson propagator in
the loop function with $\mu=800$~MeV are taken into account, and the
dashed lines show the case where only the transverse part of the
vector meson propagator in the loop function with $\mu=813$~MeV is
considered. The dash-dotted line denotes the inverse of the
$D\bar{D}^*$ potential.  } \label{fig:GTL800T813}
\end{figure}

The real and imaginary parts of the loop function $\tilde{G}$ as
functions of the total energy of the system $\sqrt{s}$ in the center
of mass frame are depicted in Fig.~\ref{fig:GTL800T813}. The solid lines denote the case where the longitudinal
part of the intermediate vector meson propagator is taken into
account, and the parameters are set to be $\mu=800$~MeV and $a=-2$.
The dashed lines show the case where only the transverse part of
the intermediate vector meson propagator in the loop function
$\tilde{G}$ is considered, and the regularization scale is set to be
$\mu=813$~MeV with the subtraction constant $a=-2$.
%

The formalism in Section \ref{sect:formalism} can be extended to study
the interaction of the $B$ meson and $\bar{B}^*$ meson by replacing the
$c$ and $\bar{c}$ quarks with $b$ and $\bar{b}$ quarks, respectively.
When the longitudinal part of the vector meson propagator in the
loop function is taken into account, the amplitude satisfies the
pole condition in Eq.~(\ref{eq:polecondition}) only in the case of
$\mu>1700$~MeV with $a=-2$ fixed. When the regularization scale $\mu$
takes the value of 1800~MeV, the pole of the amplitude is detected at
$10600.97+i0$~MeV on the complex energy plane of
$\sqrt{s}$, which is about 3~MeV lower than the $B \bar{B}^*$
threshold, and can be regarded as a bound state of the $B \bar{B}^*$
system.
If the longitudinal part of the vector meson propagator in the loop
function is excluded in the calculation, the pole of the amplitude
appears when $\mu>1974$~MeV with $a=-2$. If the regularization scale
$\mu=2000$~MeV, the pole lies at $10603.64+i0$~MeV, which is below the
$B \bar{B}^*$ threshold.
Moreover, it is worth stressing that the bound state of the
$B\bar{B}^*$ interaction has no counterpart in the 
Particle Data Group review.

\section{Summary}
\label{sect:summary3872}

The $D\bar{D}^*$ interaction is investigated in the hidden gauge
symmetry approach of the $SU(3)$ flavor subspace of the $u$, $d$ and
$c$ quark components. The one-pion exchange between the $D$ meson
and the $\bar{D}^*$ meson is analyzed precisely. Since the mass of
the $\bar{D}^*$ meson is just one pion mass larger than that of the
$D$ meson, the intermediate pion can be treated as a $real$ particle
at the $D\bar{D}^*$ threshold. Thus the diagram of the one-pion
exchange between the $D$ meson and the $\bar{D}^*$ meson is
divergent and supplies a singularity at the $D\bar{D}^*$ threshold.
However, this one-pion exchange potential becomes trivial when the
total energy of the $D\bar{D}^*$ system is far away from the
threshold, so it is neglected in this work.

A kernel of the $D\bar{D}^*$ interaction by exchanging a $\rho$ or
$\omega$ meson is derived, and then this kernel is used to solving
the Bethe-Salpeter equation in the coupled-channel unitary
approximation. In the isospin $I=0$ sector, a $D\bar{D}^*$ bound
state with a mass about $3872$~MeV is produced, which is slightly
lower than the $D\bar{D}^*$ threshold and can be regarded as a
counterpart of the $X(3872)$ particle.
This method is also extended to study the $B\bar{B}^*$ interaction
by replacing the corresponding $c$ quarks with $b$ quarks,
respectively, and a bound state is produced in the isospin $I=0$
sector, which has no counterpart in the PDG data.
It should be emphasized that the regularization scale takes
different values from the $D\bar{D}^*$ case when the subtraction
constant is fixed. The heavy quark flavor symmetry
should be considered in our future work.

Although a $D\bar{D}^*$ bound state can be generated dynamically in
the isospin $I=0$ sector, which stems from the $\rho$ and $\omega$
meson exchange due to the hidden gauge symmetry approach, the
$D\bar{D}^*$ interaction in the isospin $I=1$ sector is unfortunately zero, and thus no bound state can be generated dynamically.
This implies that other mechanisms need to be considered besides the
hidden gauge symmetry approach.

\begin{acknowledgments}
We would like to thank Han-Qing Zheng for useful discussions.
\end{acknowledgments}

\clearpage

\end{CJK*}
\end{document}